\documentclass[12pt]{iopart}
\usepackage{iopams}
\usepackage{texdraw}
\newdimen\mathindent
\newcommand{\erefX}[1]{(\ref{#1})}

\def\wPm{$\widehat P$-matrix}

\def\prodd#1#2#3{\prod\limits_{#1}^{#2}\lower3pt\hbox{${ }_{#3}$}}
\newcommand{\eqlX}[2]{\begin{equation}{#1}\label{#2}\end{equation}}

\newcommand{\eq}[1]{\begin{equation}{#1}\end{equation}}
\def\integer{\mathbb{Z}}
\def\real{\mathbb{R}}

\def\summ#1#2#3{\sum\limits_{#1}^{#2}\lower3pt\hbox{${ }_{#3}$}}

\def\prodd#1#2#3{\prod\limits_{#1}^{#2}\lower3pt\hbox{${ }_{#3}$}}

\newtheorem{theo}{Theorem}[section]

\begin{document}

\paper[Parametrization of strata in orbit space]{Tools in the
orbit space approach to the study of invariant functions: rational
parametrization of strata}

\author{G Sartori and G Valente}
\address{Dipartimento di Fisica,Universit\`a di Padova and INFN,
Sezione di Padova I--35131 Padova, Italy }

\begin{abstract}
Functions which are equivariant or invariant under the
transformations of a compact linear group $G$ acting in a
euclidean space $\real^n$, can profitably be studied as functions
defined in the orbit space of the group.  The orbit space is the
union of a finite set of strata, which are semialgebraic manifolds
formed by the $G$-orbits with the same orbit-type. In this paper,
we provide a simple recipe to obtain rational parametrizations of
the strata. Our results can be easily exploited, in many physical
contexts where the study of equivariant or invariant functions is
important, for instance in the determination of patterns of
spontaneous symmetry breaking, in the analysis of phase spaces and
structural phase transitions (Landau theory), in equivariant
bifurcation theory, in crystal field theory and in most areas
 where use is made of symmetry adapted
functions.

A physically significant example of utilization of the recipe is
given, related to spontaneous polarization in chiral biaxial
liquid crystals, where the advantages with respect to previous
heuristic approaches are shown.
\end{abstract}

\submitto{\JPA}

\eads{\mailto{gfsartori@padova.infn.it},\mailto{valente@pd.infn.it}}

\maketitle

\section{Introduction}
The determination of properties of functions which are equivariant
or invariant under the transformations of a compact linear group
(hereafter abbreviated in CLG) $G$ is often a basic problem to
solve in many physical contexts. $G$-invariant functions play an
important role, for instance, in the determination of patterns of
spontaneous symmetry breaking and structural phase transitions
(Landau theory \cite{470,620,051,557}), in equivariant bifurcation
theory (see, for instance, \cite{Chossat} and references therein),
in crystal field theory and in most areas of solid state theory.

It will not be essentially restrictive, in the following, to
assume that $G$ is a matrix subgroup of the real group O(n). In
fact, complex linear groups can always be transformed into real
linear groups through a process of realification and compact real
linear groups are equivalent to orthogonal linear groups.

An approach to the study of invariant functions, that fully
exploits invariance and possible regularity properties, takes
advantage of the fact that a $G$-invariant function $f$, defined
on $\real^n$, is a constant along each orbit of $G$ and can,
therefore, be considered as a function $\widehat f$ in the orbit
space $\real^n/G$ of the action of $G$ in $\real^n$. Geometric
invariant theory \cite{610,611} suggests how to get, in principle,
$\widehat f$ and $\real^n/G$.

The orbit space of a compact group can be realized as a connected
semi-algebraic subset ({\em i.e.}, a subset defined by algebraic
equalities and inequalities) of a Euclidean space $\real^q$. It
turns out to be formed by the union of connected semialgebraic
manifolds of different dimensions ({\em primary strata}). The
images of $G$-orbits with the same orbit--type form {\em
isotropy--type strata}, whose connected components are primary
strata.

The properties of a function defined on $\real^n/G$, may
critically depend on the geometry of this space, that has to be
known explicitly. A simple way to obtain a substantial
determination of the algebraic equations and inequalities defining
the range of an orbit map and its strata has been suggested in
\cite{020,020a,020b} (see also \cite{651,650}). These relations
can be expressed in the form of positivity and rank conditions of
a matrix $\widehat P(p)$, whose elements are polynomial functions
of $p\in\real^q$. In this way, the defining relations of the
strata are obtained in the form of algebraic equations and
inequalities.

In applications or, simply, to get a better understanding of the
geometry of the stratum (its connectivity properties, its
boundary, etc..), one often needs to solve explicitly these
relations and, sometimes, this is very difficult to do using
standard algorithms. In this paper we propose a general method to
derive the equations defining the stratification of orbit spaces
of CLG's, in the form of explicit rational parametric relations.

\vskip3truemm This paper is organized as follows. In section 2 we
summarize the results in the $P$-matrix approach to the
characterization of the orbit spaces of CLG's \cite{020,020a}. In
section 3 we present and prove our results, and in section 4, we
 illustrate them in a physically significant example, related to
spontaneous polarization in chiral biaxial liquid crystals.

\section{An overview of the geometry of linear group actions}

In this section, we shall recall, without proofs, some results
concerning invariant theory and the geometry of orbit spaces of
CLG's (see, for instance, \cite{080,710,683} and references
therein).

For our purposes, it will not be restrictive to assume that $G$ is
a matrix subgroup of the group O(n), acting linearly in the
Euclidean space $\real^{n}$ and thus defining the $G$-space
$(G,\real^n)$.

We shall denote by $g\cdot x$ the action of $g\in G$ on
$x\in\real^n$ and by $G_x$ the isotropy subgroup of $G$ at
$x\in\real^n$. The isotropy subgroups of $G$ at points of the
orbit $G\cdot x$ form the conjugacy class $(G_x)$ of $G_x$ in $G$,
which identifies the {\em orbit--type} of (the points of) $G\cdot
x$. The set of points $x\in\real^n$ (of $G$-orbits) with the same
orbit--type forms an {\em isotropy--type stratum of} $(G,\real^
n)$.

The {\em orbit space} of the action of $G$ in $\real^n$ is defined
as the quotient space $\real^ n/G$, endowed with the quotient
topology and differentiable structure. The image of a stratum of
$\real^{n}$, through the canonical projection $\pi:\
\real^n\rightarrow\real^n/G $, defines an {\em isotropy--type
stratum} of\ $\real^n/G$; the connected components of  an {\em
isotropy--type stratum} of\ $\real^n/G$ are iso--dimensional
manifolds ({\em primary strata} of $\real^n/G$), but the orbit
space is not a manifold.

Almost all the points of $\real^n/G$ belong to a unique stratum,
the {\it principal stratum}, which is a connected open dense
subset of $\real^ n/G$.  The boundary of the principal stratum is
the union of disjoint {\it singular} strata.  Every stratum
$\widehat S$ of $\real^n/G$ is open in its topological closure
$\overline{\widehat S}$.

The following partial ordering can be introduced in the set of all
the orbit--types:  $(H)<(K)$, if $H$ is conjugate with a subgroup
of $K$. If $(H_i)$ is the orbit--type of the stratum $\widehat
S_i$ of $\real^n/G$ , $i=1,2$, then $H_1<H_2$ iff $\widehat S_2$
is in the boundary of $\widehat S_1$; therefore, more peripheral
strata of $\real^n/G$ have greater orbit--type. The number of
distinct orbit--types of $G$ is finite and there is a unique
minimum orbit--type, the {\it principal orbit--type},
corresponding to the principal stratum.

The ring $\real [x]^G$ of real $G$-invariant polynomial functions
of $x$ is finitely generated \cite{286,286a,286b,Nag}. There
exists, therefore, a finite minimal collection of homogeneous
$G$-invariant polynomials $p(x) = ( p_1(x), p_{2}(x), \ldots,
p_{q}(x) )$ (minimal integrity basis for the ring of $G$-invariant
polynomials, henceforth abbreviated in MIB) such that any element
$F\in\real[x]^G$ can be expressed as a polynomial function
$\widehat F$ of $p(x)$:

\eqlX{\widehat{F}(p(x)) \,=\, F(x), \, \forall x \in
\real^{n}\,.}{e2} The number $q$ of elements of a minimal
integrity basis and their homogeneity {\em degrees} $d_i$'s are
only determined by the group $G$.

The group $G$ is said to be {\em coregular} if the elements of its
MIB's are algebraically independent. The elements of a MIB of a
non coregular group satisfy a set of algebraic identities in
$\real^n$: $\widehat F_A(p(x))\,=\,0$, $A=1,\dots$ and the
associated set of equations

\eqlX{ \widehat F_A(p)\,=\,0,\qquad A=1,\dots\,,}{e5} defines an
algebraic variety in $\real^q$, which is called the {\em variety
$\cal Z$ of the relations} (among the elements of the MIB). If $G$
is coregular, ${\cal Z}=\real^q$.

Since $G$ is a compact group, the orbits of $G$ are separated by
the elements of a MIB of $G$ (for an elementary proof see, for
instance, \cite{020a}), {\em i.e.}, at least one element of a MIB
of $G$ takes on different values at two distinct orbits. Thus, the
elements of a MIB of $G$ yield a parametrization of the points of
$\real^{n}/G$, that turns out to be also smooth, since the {\em
orbit map} $p:\, \real^{n} \longrightarrow \real^{q}$, which maps
all the points of $\real^{n}$ lying on an orbit of $G$ onto a
single point of $\real^{q}$, induces a diffeomorphism of
$\real^{n}/G$ onto a semialgebraic $q$--dimensional connected
closed subset of $\real^{q}$. Like all semialgebraic varieties
\cite{840}, $p(\real^n)$ presents a natural stratification in
connected semialgebraic sub-varieties, called {\em primary
strata}, which turn out to be the connected components of the
isotropy-type strata\footnote{A simple example of a compact
connected {\em linear} semialgebraic variety of $\real^3$ is
yielded by a polyhedron: its interior points form a unique
three--dimensional primary stratum (principal stratum), while
two-, one- and zero--dimensional primary strata are formed,
respectively, by the interior points of each face, by the interior
points of each edge, by each vertex.}.

A characterization of the image $p(\real^n)$ of the orbit space of
$G$ as a semi-algebraic variety and of its primary strata can be
easily obtained through a matrix $\widehat{P}(p)$, defined only in
terms of the $G$-invariant Euclidean scalar products between the
gradients of the elements of the MIB $\{p(x)\}$:

\begin{equation} \label{matP}
P_{ab}(x) \,=\, \sum_{i=1}^n \,\partial_i p_a(x)\,
\partial_i p_b(x) \,=\, \widehat{P}_{ab}(p(x)), \hspace{3em}
a,b = 1, \ldots, q,
\end{equation}
where in the last member, use has been made of Hilbert's theorem,
in order to express $P_{ab}(x)$ as a polynomial function of
$p_1(x), \dots ,p_q(x)$.

The following theorem \cite{020,020a,020b} (see also \cite{651})
clarifies the meaning and points out the role of the matrix
$\widehat{P}(p)$:

\begin{theo}     \label{T1}
Let $G$ be a compact subgroup of the real group O(n), $p$ the
orbit map $\real^n\rightarrow\real^q$ defined by the MIB
$(p_1(x),p_2(x),\ldots ,p_q(x))$ and $\widehat{P}(p)$ the matrix
defined in \erefX{matP}.  Then the range $p(\real^{n})$ of $p$ is
the unique semialgebraic connected subset of the variety ${\cal
Z}\subseteq \real^q$ of the relations among the elements of the
MIB where $\widehat{P}(p)$ is positive semi-definite.  The
$k$--dimensional primary strata of $p(\real^{n})$ are the
connected components of the set $\{ p \in {\cal Z} \mid
\widehat{P}(p) \geq 0, \, {\rm rank} (\widehat{P}(p)) = k \}$;
they are the images of the connected components of the
k--dimensional isotropy--type strata of $\real^{n}/G$.  In
particular, the set of interior points of $p(\real^{n})$, where
$\widehat{P}(p)$ has the maximum rank, is the image of the
principal stratum.
\end{theo}

In the following, we shall identify orbit spaces and their strata
with their images through orbit maps.

If $G$ is coregular, the $p_i$'s, with range in the semialgebraic
set $\Delta$ defined by the inequalities $\widehat P(p)>0$ (and no
equation!) provide a one-to-one parametrization of the principal
stratum of $\real^n/G$. If the stratum is singular, the $p_i$'s
are not independent parameters, being bounded also by the
equations defining the stratum. In applications or, simply, to get
a better understanding of the geometry of the stratum (its
connectivity properties, its boundary, etc. ...), one often needs
to solve explicitly these equations, which are obtained in
implicit form from the theorem recalled above. Often, this is very
difficult to do using standard algorithms. In the following
section, we propose a general method to derive the equations
defining the singular strata of the orbit space of any CLG in the
form of explicit rational parametric relations.

The mathematical apparatus just recalled offers effective tools,
for instance, in the determination of possible patterns of
spontaneous symmetry breaking, when the ground state of the system
is determined by the absolute minimum of an invariant potential.
Let us recall the basic ideas. In this context, the vector $x\in
\real^n$ is an order parameter and $G$ is the symmetry group of
the potential $\Phi(\alpha; x)$ (free energy, or Higgs potential,
for instance), expressed in terms, also, of parameters $\alpha$.
The points $x_0(\alpha)$, where the function $\phi_\alpha(x) =
\Phi(\alpha; x)$ takes on its absolute minimum, determine the
stable phase of the system, whose residual symmetry is defined by
the isotropy subgroup of $G$ at $x_0$. Owing to the $G$-invariance
of the potential, each of its stationary points is degenerate
along the $G$-orbit through it. Since the isotropy subgroups of
$G$ at points of the same orbit are conjugate in $G$, only the
orbit--type of $x_0(\alpha)$ is physically relevant. {\em
Structural phase transitions} take place when, by varying the
values of the $\alpha$'s, the point $x_0(\alpha)$ is shifted to a
different stratum.

If $\Phi(\alpha;x)$ is a sufficiently general function of the
$\alpha$'s, by varying these parameters, the point $x_0(\alpha)$
can be shifted to any stratum of $\real^n/G$. So, {\em the strata
are in a one-to-one correspondence with the symmetry phases
allowed by the $G$-invariance of the potential}. On the contrary,
extra restrictions on the form of the potential function, not
coming from G-symmetry requirements (e.g., the assumption that the
potential is a polynomial of low degree), can limit the number of
allowed structural phases for the system in its ground state.

\section{Parametrization of strata in orbit spaces of compact linear groups}

Let $G$ be a compact linear group, acting orthogonally in
$\real^n$ and $\{p\}$ be a related MIB. We shall prove that the
possibility of parametrizing the principal stratum of the orbit
space of a coregular group in terms of parameters related to the
elements of a MIB can be extended to singular strata. The result
stems from the proof of the following statement: {\em the set of
interior points of the topological closure of an isotropy--type
stratum, with orbit--type $(H)$, is diffeomorphic to the principal
stratum of the orbit space of a group space of the stabilizer of
$H$ in $G$.} If the stabilizer turns out to be coregular, the
constructive proof of this proposition will provide a one-to-one
rational parametrization of the stratum. Since, in view of
possible applications, we are mainly interested in the possibility
of getting a one-to-one parametrization of a stratum, in the
presentation of our results, stress will be laid on this aspect of
the matter.

Let $H$ be a proper isotropy subgroup of the $G$-space
$(G,\real^n)$. We shall denote by $S$ the stratum of $(G,\real^n)$
with orbit--type $(H)$ and by $\widehat S$ its image $p(S)$ in the
orbit space $\real^n/G$. Let us, also, define

\begin{equation}
V = \{x\in\real^n\mid G_x\supseteq H\},\qquad {\mathcal
V}=\{x\in\real^n \mid G_x = H\}\,. \label{1}
\end{equation}

The non trivial sets $V$  and ${\mathcal V}$ have the following
properties, which are more or less immediate consequences of their
definitions:

\begin{enumerate}
\item $V =\{x\in\real^n \mid h\cdot x=x, h\in H\}$ is a linear subspace of $\real^n$, let us call
$\nu$ its dimensions;
\item $V$ is the topological closure of $\mathcal V$: $V=\overline{\mathcal V}$;
\item ${\mathcal V}=S \cap V $ and every $G$-orbit lying in $S$ has at least a point in
${\mathcal V}$ so that $S = G\cdot {\mathcal V}$,
$\overline{S}=G\cdot V$ (where the bar denotes topological
closure) and, consequently,

\eqlX{p({\mathcal V})={\widehat S},\qquad p(V)=\overline{\widehat
S}}{pV} and

\eqlX{{\rm rank}(P(x)\big|_{x\in{\mathcal V}})={\rm dim}(\widehat
S),\qquad {\rm rank}(P(x)\big|_{x\in V\backslash{\mathcal
V}})\,<\,{\rm dim}(\widehat S).}{rango}
\end{enumerate}

A $G$-orbit of $S$ may intersect $\mathcal V$ in one or more
points. In the first case, every $G$-orbit of $S$ intersects
$\mathcal V$ in a point and, owing to item {\em (iii)}, the
coordinates of the points of ${\mathcal V}$ provide a one-to-one
rational (in effect, linear) parametrizazion of the orbits of G
lying in $S$, obtained by solving the system of linear equations
$h\cdot x=x\ \forall h\in H$. The allowed range of $x$,
$x\in{\mathcal V}$, is determined by the inequalities assuring
that rank$(P(x))=$\,dim$(\widehat S)$. The parameters $x$, with
range $\mathcal V$, would provide, in this case, a one-to-one
parametrization of the stratum, thus solving our problem. In
general, however, the intersection of a $G$-orbit, of orbit--type
$(H)$, with the stratum $S$ does not reduce to a single point. So,
a sounder analysis is required, that will be the object of the
rest of this section\footnote{Despite its being, generally, not
one-to-one, the parametrization of the orbits of a stratum by
means of the points of $V$ may be useful. It has been used, for
instance, by Kim \cite{K} to parametrize the strata of a set of
low dimensional orbit spaces of compact linear Lie groups.}.

Two distinct points, $x$ and $g\cdot x$, $g\in G$, of the same
$G$-orbit, lie in $\mathcal V$ iff $h\cdot g\cdot x = g\cdot x\
\forall h\in H$, that is, iff $g$ is in the stabilizer Stab$(H,
G)$ of $H$ in $G$.  The intersection of a $G$-orbit of $S$ with
${\mathcal V}$ is, therefore, the Stab($H$, $G$)-orbit through $x$
and Stab($H$, $G$) is the largest subgroup of $G$ leaving $V$
invariant.

In the group space (Stab($H$, $G$),V), the isotropy subgroup at a
point of general position is $H$. Therefore, the principal stratum
$\Sigma$ satisfies the following relation:

\eqlX{V\supset \Sigma\supseteq {\mathcal V}.}{ins} It has to be
noted that $\Sigma$ could contain $\mathcal V$ in a strict sense,
since, at points of $V\backslash{\mathcal V}$, the conjugacy class
in $G$ of the isotropy subgroup of Stab$(H,G)$ could be smaller
than the conjugacy class in $G$ of the isotropy subgroup of $G$.

Now, $H$ is an invariant subgroup of Stab$(H,G)$ and a subgroup of
all the isotropy subgroups of (Stab$(H,G),V)$. Thus, the action of
Stab$(H,G)$ in $V$ defines a linear group $K$ (and the group space
$(K,V)$), isomorphic to the quotient group Stab($H$, $G$)/$H$,
through the relation

\eqlX{(sH)\cdot v=s\cdot v,\qquad s\in {\rm Stab}(H,G),\ sH\in
{\rm Stab}(H,G)/H.}{K} So, we can conclude that {\em the
intersection of a $G$-orbit of $S$ with ${\mathcal V}$ is an orbit
of $(K,V)$}. Obviously, the orbit spaces $V/K$ and $V/{\rm
Stab}(H,G)$ are isomorphic and can be identified.

We can rephrase the result just obtained, by claiming that the
points of $\widehat S$ are in a one-to-one correspondence with the
points of a, possibly proper, subset $\widehat{\mathcal V}$ of the
principal stratum $\widehat\Sigma$ of the orbit space $V/K$ and
the following relations hold true:

\eqlX{\overline{\widehat S}\,=\,p(V)\,\supset\,
p(\Sigma)\,\supseteq\, p({\mathcal V})\, =\, \widehat S.}{pS}
Since, as stressed, we know how to parametrize a principal
stratum, our problem is reduced to the determination of
$\widehat{\mathcal V}$. We shall show that this can be easily
done, making use of \erefX{rango}.

To make simpler the solution of the problem, let us assume that an
orthonormal basis has been introduced in $\real^n$ such that the
first $\nu$ elements of the basis yield a basis for the vector
space $V$. Then, if we denote by $V_\bot$ the orthogonal
complement of $V$ in $\real^n$, the subspace $V_\bot$ is invariant
under ($K$ and) $H$, owing to the orthogonality of the
transformations of $G$. Since $V$ contains all the $H$-invariant
vectors of $\real^n$, there is no non trivial $H$-invariant vector
in $V_\bot$.

To attain our goal, we shall start from a convenient
parametrization of the principal stratum $\widehat \Sigma$ of
$V/K$ in terms of $l$ real parameters $\lambda$, related to a MIB
$(\lambda_1(v),\dots ,\lambda_l(v))$ of the ring of polynomial
$K$-invariant functions of $v \in V$. This parametrization will be
global and one-to-one, if $K$ turns out to be coregular. In this
case, the range of $\lambda$ has to be restricted to the
positivity region of the $\widehat P$-matrix
$\widehat\Lambda(\lambda)$, associated with the MIB $\{\lambda\}$:

\eqlX{\widehat\Lambda_{\alpha\beta}(\lambda(v)) =
\Lambda_{\alpha\beta}(v) = \sum_{i=1}^\nu\,
\partial_i \lambda_\alpha(x_1,\dots ,x_\nu)\, \partial_i \lambda_\beta(x_1,\dots ,x_\nu).}{4}
As recalled in the previous section, in fact, the orbit space
$V/K$ and its principal stratum $\widehat \Sigma$ can be
identified, respectively, with the semialgebraic sets $\lambda(V)$
and $\lambda(\Sigma)$:

\eqlX{\widehat\Sigma \,=\,
\lambda(\Sigma)\,=\,\{\lambda\in\real^l\mid \widehat
\Lambda(\lambda)>0\},\qquad V/K= \overline{\widehat
\Sigma}\,=\,\lambda(V)\,=\,\{\lambda\in\real^l\mid
\widehat\Lambda(\lambda)\ge 0\},}{20} the second set being the
closure of the first one. Moreover, the definition of
$\widehat{\mathcal V}$ and equation \erefX{ins} imply

\eqlX{\widehat{\mathcal V}=\lambda({\mathcal
V})\subseteq\lambda(\Sigma)\,=\,\widehat\Sigma.}{inshat}

If $K$ is not coregular, only a local one-to-one parametrization
can be obtained for $\widehat \Sigma$, by eliminating redundant
elements in the set of parameters $(\lambda_1,\dots ,\lambda_l)$,
through the solution of the algebraic relation(s) among the
elements of the MIB $\{\lambda\}$, and imposing convenient
semi-positivity and rank conditions on the matrix
$\widehat\Lambda(\lambda)$.

A one-to-one local or global parametrization of $\widehat \Sigma$
yields, obviously, also a local or global one-to-one
parametrization of $\widehat {\mathcal V}$, provided that
additional restrictions are imposed on the range of $\lambda$,
whenever $\widehat {\mathcal V}$ is a proper subset of
$\widehat\Sigma$. In this case, the correct bounds can be obtained
in the following way (we shall only consider the case of a
coregular $K$, the extension of the results to non coregular $K$'s
is straightforward, but, as just stressed, may lead to a loss of
globality).

When $x$ spans $V$, the elements of the MIB
$(p_1(x),\dots,p_q(x))$ of $G$ define a set of $K$-invariant
polynomial functions of $v=(x_1,\dots ,x_\nu)\in V$. Therefore, by
the Hilbert theorem recalled in the previous section, they can be
expressed as polynomial functions of $\lambda$, that is,

\eqlX{p\big|_V=\phi\circ\lambda.}{phi} Let us remark that, in our
assumptions, there are no relations among the elements of the MIB
$\{\lambda\}$ and, consequently, possible relations
$F_\alpha(p)=0$ among the $p_i$'s are identically satisfied for
$p=\phi(\lambda)$.

From \erefX{phi} and \erefX{ins}, one immediately obtains

\eqlX{\overline{\widehat S}\,=\,p(V)\,=\,\phi(V/K)\,\supset\,
\phi(\widehat\Sigma)\,\supseteq\,\phi(\widehat{\mathcal
V})\,=\,p({\mathcal V})=\widehat S}{pins} and, since
$\widehat\Sigma$, being a principal stratum, is the set of
interior points of $V/K$ and is connected, $\phi(\widehat\Sigma)$
will coincide with the set of interior points of the closure
$\overline{\widehat S}$ of $\widehat S$ and will be connected.
This set does not coincide with $\widehat S$ if
$\overline{\widehat S}$ contains, in its interior, points
representing bordering strata of $\widehat S$. This certainly
happens if the set $\widehat S$ is not connected and, presumebly,
also if it is not multiply connected\footnote{To our knowledge, no
general proof exists even of simple connectivity of the principal
stratum of a coregular compact group; however, principal strata of
coregular low dimensional ($D\le 4$) orbit spaces can be checked
to be multiply connected \cite{685,fini}.}.

The identification of points $\lambda\in\widehat \Sigma$, if any,
whose image $\phi(\lambda)\not\in\widehat S$, can be obtained in
the following way.

Let $x=v\oplus v_\bot$ yield the decomposition of $x\in\real^n$ in
its vector components $v\in V$ and $v_\bot\in V_\bot$:
$v=(x_1,\dots ,x_\nu)$, $v_\bot=(x_{\nu+1},\dots ,x_n)$. Then, a
$G$-invariant polynomial $f(x)$, can be thought of as a polynomial
in $v$ and $v_\bot$ and it is easy to realize that it cannot
contain linear terms in $v_\bot$, being $v$ invariant under $H$.
As a consequence,

\eqlX{\partial_i f(x)=0,\ {\rm for\ }x\in V\ {\rm and\
}i=\nu+1,\dots ,n.}{3} Now, starting from the very definition of
$P(x)$ (see \erefX{matP}), for every $x=v\in V$ we obtain, using
\erefX{3} and the identity $p(v)=\phi(\lambda(v))$,
\eqlX{\begin{array}{rcl}
\widehat P_{ab}(p(v))&=&\sum_{i=1}^\nu\,\partial_i p_a(v)\,\partial_i p_b(v)\\
&&\\
&=& \left.
\sum_{\alpha,\beta=1}^l\,\partial_\alpha\phi_a(\lambda)\,
\partial_\beta\phi_b(\lambda)
\right|_{\lambda=\lambda(v)}\,\sum_{i=1}^\nu\,\partial_i
\lambda_\alpha(v)\,
\partial_i\lambda_\beta(v)\\
&&\\
&=& \left.
\left(J(\lambda)\,\widehat\Lambda(\lambda)\,J(\lambda)^{\rm
T}\right|_{\lambda=\lambda(v)}\right)_{ab} ,
\end{array}}{5}
where, the superscript $T$ denotes transposition and $J(\lambda)$
is the Jacobian matrix of the transformation $p=\phi(\lambda)$:

\eqlX{J_{a\alpha}(\lambda)=\partial_a\phi_\alpha (\lambda),\qquad
a=1,\dots ,q,\ \ \alpha=1,\dots ,l.}{7} So, we can conclude that,
for all $\lambda\in V/K$,

\eqlX{\widehat P(\phi(\lambda))=
J(\lambda)\,\widehat\Lambda(\lambda)\,J(\lambda)^{\rm T}.}{6}

Equation \erefX{6} leads to an easy calculation of the points
$\lambda\in \widehat{\mathcal V}$, that is, of the points
$\lambda\in \widehat\Sigma$ whose image $p=\phi(\lambda)$ is in
$\widehat S$. In fact, from \erefX{rango}, these points are
characterized by the conditions $\lambda\in\widehat\Sigma$ and
rank($\widehat P(\phi(\lambda)))=l$. Since, for
$\lambda\in\widehat\Sigma$,  the matrix $\widehat
\Lambda(\lambda)$ is positive definite with rank $l$, we can
conclude that the range of $\lambda$ assuring the location of the
point $p=\phi(\lambda)$ in $\widehat S$ is the semialgebraic
subset $\Delta$ of $\real^l$, determined by the following
inequalities:

\eqlX{\Delta=\{\lambda\in\real^l\mid \widehat{\Lambda}(\lambda)>0
\ {\rm and}\ {\rm rank}(J(\lambda))=l\}.}{8} For
$\lambda\in\Delta$, the relation $p=\phi(\lambda)$ yields a global
rational parametrization for $\widehat S$.

It will be worthwhile to stress that the parametrization we have
suggested is, in some way, canonical: the unique arbitrariness in
the choice of the parameters is related to the choice of the
MIB's.

An important byproduct of the result just proved is a simple test
of the connection of $\widehat S$, that could be difficult or
impossible to read directly from the equations of the stratum in
implicit form. In fact, the condition that the boundary of
$\overline{\widehat S}$ coincides with the boundary of $\widehat
S$ is equivalent to the condition rank($J(\lambda))=l$ for all
$\lambda\in\widehat \Sigma$.

If $K$ is not coregular, there are polynomial relations
$F_A(\lambda)=0$ among the elements of the MIB $(\lambda_1,\dots
,\lambda_l)$ and $l>$dim$(\widehat S)$. As already stressed, to
obtain a one-to-one parametrization of the points of $\widehat S$
by means of the $\lambda_\alpha$'s, one has to eliminate the
redundant parameters by solving the equations $F_A(\lambda)=0$.
This may be feasible, but only locally and the resulting
parametrization will not be global and possibly not rational.

\section{An Example}
In this section we shall show how the parametrization technique
works in a simple example. The notations will be the same defined
in the previous sections.

We shall consider the orthogonal linear group $G$ defined by the
action of the real group O(3) in the real eight--dimensional space
spanned by the independent components $x_1,\dots ,x_5$ of a
symmetric and traceless tensor $Q$ and the three components
$x_6,x_7,x_8$ of a polar vector $\bi{P}$. To be specific, if $O$
is a generic $3\times 3$ real orthogonal matrix, the
transformation rules of the $x_i$ are obtained from the following
relations:

\begin{equation}
Q = \frac{1}{\sqrt{2}} \left(
\begin{array}{ccc}
-{\displaystyle\frac{2}{\sqrt{3}}} x_1&  x_3 & x_4 \\
x_3                       & {\displaystyle\frac{x_1}{\sqrt{3}}} - x_2 & x_5 \\
x_4                       &  x_5                & {\displaystyle\frac{x_1}{\sqrt{3}}} + x_2 \\
\end{array}
\right)
\end{equation}

\begin{equation}
\begin{array}{rcll}
Q'_{\alpha\beta}&=& \sum_{\gamma\,\delta=1}^3 O_{\alpha\gamma}\,
                                    Q_{\gamma\delta}\,  O_{\beta\delta}& \\
                                    &&&\alpha,\beta=1,2,3\,.    \\
P'_{\alpha}&=& \sum_{\beta=1}^3 O_{\alpha\beta}\, P_{\beta}\;,
\end{array}\label{mu2}
\end{equation}
It is worthwhile to remark that the reflection $I=\mbox{\rm diag}
(-1,-1,-1)\in$ O(3) reverses only the signs of the last three
coordinates $(x_6,x_7,x_8)$ and, due to the symmetry of the tensor
$Q$, each $G$-orbit contains points where $Q$ takes on a diagonal
form ($x_3=x_4=x_5=0)$. This remark makes much easier the
determination of ``typical points" in strata, from which the orbit
type of the stratum can be identified \cite{Li,K}.

The real orthogonal linear group $G$, just defined, is coregular
(see \cite{711}) and its isotropy subgroup at a generic point of
$\real^8$ is trivial. As a consequence, the principal orbits are
three--dimensional manifolds, the principal stratum has dimensions
five and there are five independent elements in a MIB (see, for
instance \cite{683}), which can be chosen in the following way:

\begin{eqnarray}
p_1 &=&\Tr\, Q^{2} + \bi{P}\cdot
\bi{P} \nonumber\\
    &=& \sum_{j=1}^{8} {x_j}^{2}\,,\nonumber\\
&&\nonumber\\
p_2 &=& \bi{P}\cdot \bi{P} \nonumber\\
    &=&  \sum_{j=6}^{8} {x_j}^{2}\,,\nonumber\\
&&\nonumber\\
p_3 &=& 6 \sqrt{2} \,\,\Tr\, Q^{3}\nonumber\\
    &=& -2\, {\sqrt{3}}\, {{x_1}}^3 + 6\, {\sqrt{3}}\, {x_1}\, {{x_2}}^2
    - 3\, {\sqrt{3}}\, {x_1}\, {{x_3}}^2 - 9\,  {x_2}\, {{x_3}}^2
      \label{BaseInt}\\
  && - 3\, {\sqrt{3}}\, {x_1}\, {{x_4}}^2+ 9\, {x_2}\, {{x_4}}^2+18\,
{x_3}\, {x_4}\, {x_5} +   6\, {\sqrt{3}}\, {x_1}\, {{x_5}}^2 \,,\nonumber\\
&& \nonumber\\
p_4 &=& 3 \sqrt{2} \,\sum_{\alpha\,\beta}P_\alpha Q_{\alpha\,\beta}
P_{\beta}\nonumber\\
 &=& -2\,{\sqrt{3}}\, {x_1}\, {{x_6}}^2  +
  6\, {x_3}\, {x_6}\, {x_7} + {\sqrt{3}}\, {x_1}\, {{x_7}}^2 -
  3\, {x_2}\, {{x_7}}^2  \nonumber\\
  && + 6\, {x_4}\, {x_6}\, {x_8}+ 6\, {x_5}\, {x_7}\, {x_8}+ \sqrt{3}\,
{x_1}\, {{x_8}}^2+3\, {x_2}\, {{x_8}}^2\,,\nonumber\\
&&\nonumber\\
p_5 &=& 6 \,\sum_{\alpha\,\beta}P_\alpha {Q^{2}}_{\alpha\,\beta}
P_{\beta}\nonumber\\
    &=& 4\,{{x_1}}^2\, {{x_6}}^2+ 3\, {{x_3}}^2\, {{x_6}}^2 +
  3\, {{x_4}}^2\, {{x_6}}^2-2\, {\sqrt{3}}\, {x_1}\, {x_3}\,
{x_6}\,{x_7}
    \nonumber\\
  &&- 6 \,{x_2}\,{x_3}\,{x_6}\,{x_7}+ 6 \,{x_4}\,{x_5}\,{x_6}\,{x_7} +
   {{x_1}}^2\, {{x_7}}^2 -
  2\, {\sqrt{3}}\, {x_1}\, {x_2}\, {{x_7}}^2 \nonumber\\
  &&+ 3\, {{x_2}}^2\, {{x_7}}^2 +
  3\, {{x_3}}^2\, {{x_7}}^2 + 3\, {{x_5}}^2\, {{x_7}}^2 -
  2\, {\sqrt{3}}\, {x_1}\, {x_4}\, {x_6}\, {x_8}   \nonumber\\
  && +  6\, {x_2}\, {x_4}\, {x_6}\, {x_8}+6\, {x_3}\, {x_5}\, {x_6}\,
  {x_8} +     6\, {x_3}\, {x_4}\, {x_7}\, {x_8} +
  4\, {\sqrt{3}}\, {x_1}\, {x_5}\, {x_7}\, {x_8}   \nonumber\\
  &&+   {{x_1}}^2\, {{x_8}}^2
  +2\,{\sqrt{3}}\, {x_1}\, {x_2}\, {{x_8}^2}+3\, {{x_2}}^2\, {{x_8}}^2 +
  3\, {{x_4}}^2\, {{x_8}}^2   +3\, {{x_5}}^2\, {{x_8}}^2\;.\nonumber
  \end{eqnarray}
The corresponding $\widehat P $-matrix elements can be easily
calculated from their definition \erefX{matP}:
\begin{equation}
\begin{array}{rcl}
\widehat{P}_{1\,a} &=& 2 d_a p_a\;, \hspace{3em} 1 \leq a \leq 5 \\
\widehat{P}_{2\,2} &=& 4 p_2 \\
\widehat{P}_{2\,3} &=& 0 \\
\widehat{P}_{2\,4} &=& 4 p_4 \\
\widehat{P}_{2\,5} &=& 4 p_5 \\
\widehat{P}_{3\,3} &=& 108 \left(p_1 -p_2\right)^{2} \\
\widehat{P}_{3\,4} &=& 18\,\left( - 2\,p_1\,p_2+ 2 {p_2}^2+ p_5 \right) \\
\widehat{P}_{3\,5} &=& 12\,\left( \,p_2\,p_3+ p_1 \,p_4 -p_2 \, p_4 \right) \\
\widehat{P}_{4\,4} &=& 12 \,\left({p_2}^2 + p_5 \right)\\
\widehat{P}_{4\,5} &=& 4\,\left( \,p_2\,p_3+3\, p_1 \,p_4 -p_2 \, p_4 \right) \\
\widehat{P}_{5\,5} &=& \frac{4}{3} \left(p_3 \, p_4 + {p_4}^2 + 9\, p_1 p_5 \right)\;,
\end{array}
\end{equation}
where the ordered set $(d_1,d_2,d_3,d_4,d_5)=(2,2,3,3,4)$
specifies the degrees of the polynomials of the MIB.

The determinant of the matrix $\widehat P(p)$ factorizes and
only one of the two real irreducible factors,
that we shall call $A(p)$, turns out to be {\em active} \cite{682}, that is, to be
relevant in the determination of the boundary of the orbit space $\real^8/G$:

\begin{equation} \label{attivo}
\begin{array}{rcl}
A(p)&=& 3\,{{p_2}}^3\,{{p_3}}^2 + 18\,{p_1}\,{{p_2}}^2\,{p_3}\,{p_4} -
  18\,{{p_2}}^3\,{p_3}\,{p_4} +
  27\,{{p_1}}^2\,{p_2}\,{{p_4}}^2 \\
  && -
  54\,{p_1}\,{{p_2}}^2\,{{p_4}}^2 + 27\,{{p_2}}^3\,{{p_4}}^2 +
  {p_3}\,{{p_4}}^3 - 9\,{p_2}\,{p_3}\,{p_4}\,{p_5}  \\
  &&
  -9\,{p_1}\,{{p_4}}^2\,{p_5} + 9\,{p_2}\,{{p_4}}^2\,{p_5} -
  27\,{p_1}\,{p_2}\,{{p_5}}^2 + 27\,{{p_2}}^2\,{{p_5}}^2 +
  9\,{{p_5}}^3\;.\\
\end{array}
\end{equation}

\begin{center}
\begin{figure}
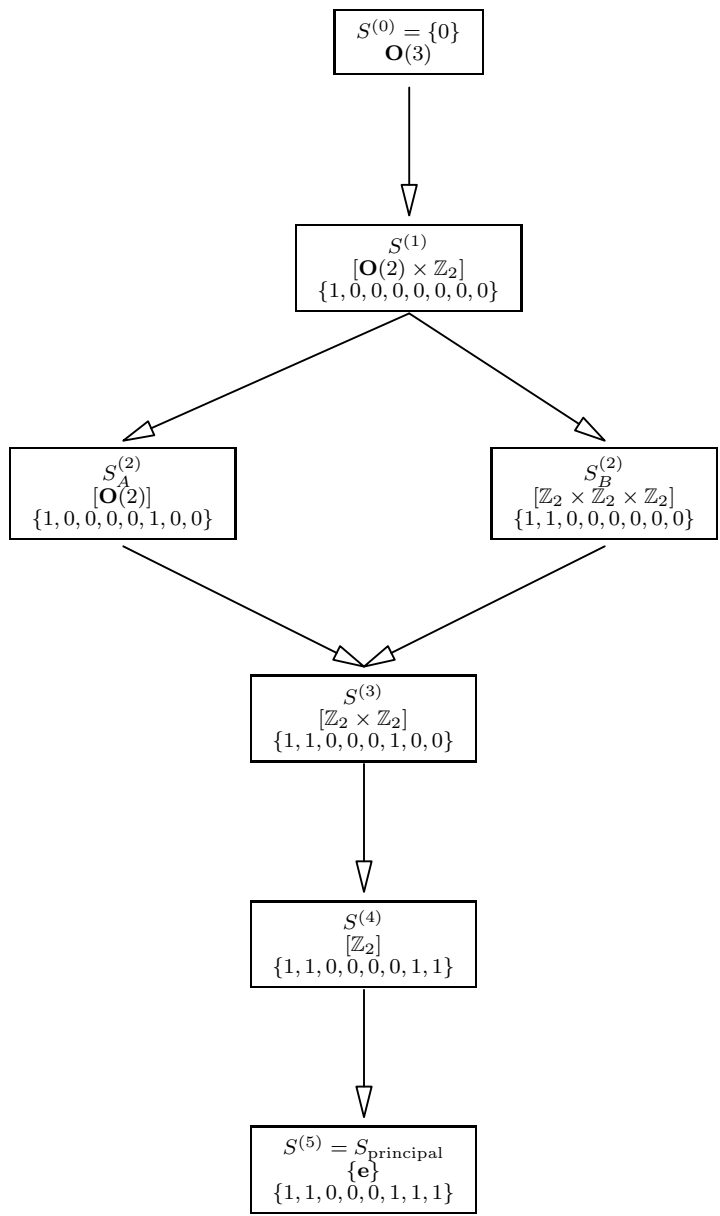

\caption{\label{F1} Possible phase transitions between bordering strata are
connected by continuous sequences of one or more arrows. For each
possible structural phase (isotropy type stratum), a ``typical''
point in the order parameter space and the corresponding isotropy
subgroup of O$(3)$ are indicated. }
\begin{center}
\begin{texdraw}

\drawdim mm

\move(80 160) \textref h:C v:C \htext{\ }

\move(85 150) \textref h:C v:C
\htext{\framebox{\scriptsize$\def\arraystretch{0.8}\begin{array}{c}
S^{(0)}=\{0\}\\{\rm\bf O}(3)\end{array}$}}

\move(85 144) \avec(85 127)

\textref h:C v:C \htext(85
120){\framebox{\scriptsize$\def\arraystretch{0.8}\begin{array}{c}
S^{(1)}\\ {[}
{\rm \bf O}(2) \times {\mathbb{ Z}}_2 {]} \\
\{ 1,0,0,0,0,0,0,0\}\end{array}$}}

\move(85 114) \avec(47 97) \move(85 114) \avec(111 97)

\textref h:C v:C \htext(47
90){\framebox{\scriptsize$\def\arraystretch{0.8}\begin{array}{c}
S^{(2)}_{A}\\ {[}
{\rm \bf O}(2) {]} \\
\{ 1,0,0,0,0,1,0,0\} \end{array}$}}

\textref h:C v:C \htext(111
90){\framebox{\scriptsize$\def\arraystretch{0.8}\begin{array}{c}
S^{(2)}_{B}\\ {[}
{\mathbb{ Z}}_2 \times {\mathbb{ Z}}_2 \times {\mathbb{ Z}}_2{]}\\
\{ 1,1,0,0,0,0,0,0\}\end{array}$}}

\move(47 83) \avec(79 67) \move(111 83) \avec(79 67)

\textref h:C v:C \htext(79
60){\scriptsize\framebox{$\def\arraystretch{0.8}\begin{array}{c}
S^{(3)}\\ {[}
{\mathbb{ Z}}_2 \times {\mathbb{ Z}}_2 {]}\\
\{ 1, 1, 0,0,0,1,0,0 \} \\
\end{array}$}}

\move(79 54) \avec(79 37)

\textref h:C v:C \htext(79
30){\scriptsize\framebox{$\def\arraystretch{0.8}\begin{array}{c}
S^{(4)}\\ {[}
{\mathbb{ Z}}_2  {]} \\
\{ 1, 1, 0,0,0,0,1,1 \}\end{array}$}}

\move(79 24) \avec(79 7)

\textref h:C v:C \htext(79 0){\scriptsize\framebox{$\def\arraystretch{0.8}\begin{array}{c}  S^{(5)}=S_{\rm principal} \\
 \{{\bf e}\}\\
\{ 1, 1, 0,0,0,1,1,1 \}\end{array}$}}

\end{texdraw}
\end{center}
\end{figure}
\end{center}

The relations defining strata of dimension $<4$ in the orbit space
of (O(3),$\real^8$) are summarized in \tref{tabsing}. The isotropy
subgroup lattice, with the possible phase transitions between
bordering strata is shown in \fref{F1}.

\subsection{Parametrization of the strata}
The relations assuring that $\widehat P(p)\ge 0$ and has rank 4,
define a unique four--dimensional stratum $\widehat S^{(4)}$ in
the orbit space. Using well known matrix theory results, these
conditions could be explicitly written, for instance, in the form
$A(p)=0$ and $M_i(p)>0$, $i=1,\dots ,4$, where $M_i$ is the sum of
the principal minors of order $i$ of the matrix $\widehat P(p)$: a
cumbersome set of conditions that it is not worthwhile to write
down explicitly.

Even in this simple example one immediately realizes that the
choice of a minimal set of explicit algebraic relations providing
a cylindrical decomposition \cite{Coste} for the semi-algebraic
subset $\widehat S^{(4)}$ of $\real^{5}$ would be a really hard
task (for the more peripheral strata, instead, the problem is much
easier to solve). An immediate application of the results proved
in the previous section, on the contrary, leads to a simple
rational global parametrization of each stratum, as shown below.

\subsubsection{Stratum $\widehat S^{(4)}$.}
A ``typical point" of the stratum is $x_{\rm
t}=(1,1,0,0,0,0,1,1)$. The corresponding isotropy subgroup $H$ of
$G$ is the $\integer_2$ group generated by the reflection
representing the element diag$(-1,1,1)\in$ O(3) in $\real^8$:
\[\mbox{\rm diag}(1,1,-1,-1,1,-1,1,1)\in G.\]

The vector space $V$ formed by the $H$-invariant vectors of
$\real^8$ turns out to be five--dimensional:
\[V = \left\{x\in \real^8 \;|\;x_3=x_4=x_6=0\right\}.\]

The elements of  O(3), corresponding to elements of $\mbox{\rm
Stab }(H,G)$, are block-diagonal matrices of the form diag$(\pm
1,O)$, with $O\in $ O(2). Therefore, $\mbox{\rm Stab }(H,G)$ is
isomorphic to a group $\integer_2 \times$O(2) and the quotient
group $K=\mbox{\rm Stab }(H,G)/H$ is the representation in $V$
(induced by the representation $G$ of O(3)) of the O$_1(2)$
subgroup of O(3), formed by the rotations around the first axis.

Using coordinates $v=(x_1, x_2,x_5,x_7,x_8)$ for a vector $v\in
V$, the elements of $K$ turn out to be block diagonal matrices
leaving invariant the following subspaces $V^{A}$, $1 \leq A \leq
3$:
\begin{eqnarray}
V^{1}& = &\{(x_1,0,0,0,0) \in V | x_1 \in \real\}\,, \nonumber \\
V^{2}& = &\{(0,x_2,x_5,0,0) \in V | x_2\,,x_5  \in \real\} \,,\nonumber \\
V^{3}& = &\{(0,0,0,x_7,x_8) \in V | x_7\,,x_8 \in \real\}\,.
\end{eqnarray}

A proper rotation $r(\phi)\in$ O$_1(2)$ of an angle $\phi$ and the
reflection diag(-1,1)$\in {\rm O}_1(2)$ are represented in $V$ by
the $5\times 5$ matrices $1\oplus r(-2\phi)\oplus r(\phi)$
and ${\rm diag}(1,-1,1,-1,1)$.

Noting that the complex variables $z_1=x_2+\rmi x_5$ and
$z_2=x_7+\rmi x_8$ transform into $\exp(-2\rmi \phi)\,z_1$ and,
respectively, $\exp(\rmi \phi)\,z_2$ under a rotation
 and into $z_1^*$ and, respectively, $-z_2^*$ under a
reflection, it is easy to realize that a possible MIB for $(K,
V)$ is the following:
\begin{eqnarray}
\lambda_1 &=& x_1 \,,\nonumber\\
\lambda_2 &=&|z_1|^2\,=\, x_2^2 + x_5^2\,,\nonumber\\
\lambda_3 &=& |z_2|^2\,=\,x_7^2 + x_8^2\,,\nonumber\\
\lambda_4 &=& 2 \, \mbox{\rm Re} (z_1 {z_2}^2)\,=\,2
\left( x_2 {x_7}^2 -2 x_5 x_7 x_8 - x_2 {x_8}^2 \right)\;.
\end{eqnarray}

It is, now, easy to express the $p$'s in term of the $\lambda$'s, $p=\phi(\lambda)$, and to check that
the following expressions, obtained in this way, identically satisfy the equation
$A(p(\lambda))=0$ (see (\ref{attivo})) entering the definition of the stratum $\widehat S^{(4)}$:
\begin{eqnarray} \label{parametrizzazione}
\phi_1(\lambda) &=& {\lambda_1}^2 + \lambda_2 + \lambda_3 \,,\nonumber\\
\phi_2(\lambda) &=& \lambda_3\,,\nonumber\\
\phi_3(\lambda) &=& -2 \, \sqrt{3}\,\lambda_1\,\left( {\lambda_1}^2 - 3 \lambda_2
\right)\,,\nonumber\\
\phi_4(\lambda) &=& \frac{\sqrt{3}}{2} \,\left( 2 \lambda_1 \lambda_3 - \sqrt{3}
\lambda_4 \right)\,, \nonumber\\
\phi_5(\lambda) &=& {\lambda_1}^{2} \lambda_3 + 3 \lambda_2
\lambda_3 - \sqrt{3} \lambda_1 \lambda_4.
\end{eqnarray}

As explained in the previous section, since the group $K$ is
coregular, the range  $\Delta$ for $\lambda$ is the region where
the \wPm\/ $\widehat{\Lambda}(\lambda)$ associated to the MIB
$\{\lambda\}$ is positive definite and the rank of the Jacobian
matrix $J(\lambda)$ of the transformation $\phi(\lambda)$ is
maximum (=4). The matrices $\widehat\Lambda(\lambda)$ and
$J(\lambda)$ are easily calculated to be

\begin{equation} \label{lamdilam}
\widehat{\Lambda}(\lambda) = \left(
\begin{array}{cccc}
1 & 0 & 0 & 0 \\
0 & 4 \lambda_2 & 0 & 2 \lambda_4 \\
0 & 0 & 4 \lambda_3 & 4 \lambda_4 \\
0 & 2 \lambda_4 & 4 \lambda_4 & 4\,\left( {\lambda_3}^2 + 4
\lambda_2 \lambda_3 \right)
\end{array}
\right)
\end{equation}
and

\begin{equation}
J(\lambda) = \left(
\begin{array}{cccc}
2 \lambda_1 & 1 & 1 & 0 \\
0 & 0 & 1 & 0 \\
- 6\,\sqrt{3} \,\left({\lambda_1}^2 - \lambda_2 \right) &
        6\,\sqrt{3}\, \lambda_1 & 0 & 0 \\
\sqrt{3} \lambda_3 & 0 & - \sqrt{3} \lambda_1 & -3/2 \\
2 \lambda_1 \lambda_3 - \sqrt{3} \lambda_4 & 3 \lambda_3 &
{\lambda_1}^2 + 3 \lambda_2 & - \sqrt{3} \lambda_1
\end{array}
\right)\;.
\end{equation}
The conditions assuring the positivity of $\widehat{\Lambda}(\lambda)>0$ can be written
in the form

\begin{equation}
\Delta = \left \{(\lambda_1,\lambda_2, \lambda_3, \lambda_4) \in
\real^4 \;|\; \lambda_2 >0 \; \mbox{and} \; \lambda_3 >0
\;\mbox{and}\; \lambda_4^2 < 4 \lambda_2 {\lambda_3}^2 \right \}
\end{equation}
and the rank of $J(\lambda)$ turns out to equal 4 for all
$\lambda \in \Delta$. So we can conclude that the stratum is
connected, a piece of information that would be difficult to
derive directly from the relations defining the stratum in
implicit form ({\em i.e.} $A(p)=0$ and $M_i(p)>0$).

The parametrization obtained for the sub-peripheral stratum
$\widehat S^{(4)}$ turns out to be useful, also, because bordering
values of $\lambda$ immediately determine bordering values of $p$,
corresponding to more peripheral strata of $\real^8/G$. The
stratification of $ V/K$ is summarized in \tref{tab2}.
We would like to remark, however, that there is not a one-to-one
correspondence between singular strata of $(K,V)$ corresponding to
bordering values of $\lambda$ and singular strata of $\real^n/G$,
corresponding to the associated values of $p(\lambda)$. In fact,
as was already noted, since the action of the stabilizer on the
subspace $V^{1}$ is trivial, the invariant $\lambda_1$ has degree
$1$, so it does not participate in the conditions defining the
stratification of $ \left(K,V \right)$. That is why the
parametrization procedure has to be applied {\em stratum per
stratum}.

\subsubsection{Stratum $\widehat S^{(3)}$.}
A typical point in this stratum is $x_{\rm t}=(1,1,0,0,0,1,0,0)$
and the isotropy subgroup at $x_{\rm t}$ is the $\integer_2 \times
\integer_2$ group represented by the O(3) elements $\mbox{\rm
diag}( 1, \pm1, \pm 1)$. The vector space left invariant by $H$
reduces to the three--dimensional space $V$:

\begin{equation}
V = \left\{x \in \real^8 \;|\;x_3=x_4=x_5=x_7=x_8=0\right\}\;.
\end{equation}
The linear group $K$, acting in $V$, is a
$\integer_2\times\integer_2$ group, generated by the matrices
$\mbox{\rm diag}(1,-1,1)$ and $\mbox{\rm diag}(1,1,-1)$. A MIB for
$\mbox{\rm Stab }(K,V)$ is
\begin{equation}
(\lambda_1, \lambda_2, \lambda_3)  =( x_1 ,
  {x_2}^2,
  {x_6}^2 )\;.
\end{equation}
 The associated matrix $\widehat\Lambda(\lambda)$ turns out to be
\eq{\widehat\Lambda(\lambda)=\mbox{\rm diag}(1, 4 \lambda_2, 4 \lambda_3)\,.}
Parametric equations,
$p=\phi(\lambda)$, for the stratum can be written in the following
form:
\begin{eqnarray} \label{parametrizzazione1}
\phi_1(\lambda) &=& {\lambda_1}^2+\lambda_2 + \lambda_3 \,,\nonumber\\
\phi_2(\lambda) &=& \lambda_3\,,\nonumber\\
\phi_3(\lambda) &=& -2\,\sqrt{3} \lambda_1 \left( {\lambda_1}^2 - 3 \lambda_2\right)\,,\nonumber\\
\phi_4(\lambda) &=& -2\,\sqrt{3} \lambda_1 \lambda_3 \,, \nonumber\\
\phi_5(\lambda) &=&  4 {\lambda_1}^2 \lambda_3\,,
\end{eqnarray}
for $\lambda$ in the range $\Delta = \left \{(\lambda_1,\lambda_2,
\lambda_3) \in \real^3 \;|\; \lambda_3 >0  \;\mbox{\rm and} \;
\lambda_2 >0 \right \}$.

The rank of the jacobian matrix $J(\lambda)$  turns out to
equal $3$ for all $\lambda \in
\Delta$, so that the stratum is connected.

\subsubsection{Stratum $S^{(2)}_A$.}
A typical point is $x_{\rm t}=(1,0,0,0,0,1,0,0)$ and the isotropy
subgroup $H$ of $G$ at $x_{\rm t}$ is the subgroup formed by the
rotations (proper and improper) around the first axis, represented
by block-diagonal matrices of the form diag$(1,O)$, with $O\in $
O(2). The vector space $V$ left invariant by $H$ is
two--dimensional:

\begin{displaymath}
 V = \left\{ x \in \real^8 \;
\mid\;x_2=x_3=x_4=x_5=x_7=x_8=0\right\}\,.
\end{displaymath}
$K$ is the $\integer_2$ group generated by the matrix
diag$(1,-1)$. A MIB for $(K,V)$ is the following:
\begin{equation}
(\lambda_1, \lambda_2) =(x_1 , {x_6}^2) \,.
\end{equation}
The associated matrix $\widehat\Lambda(\lambda)$ turns out to be
\eq{\widehat\Lambda(\lambda)=\mbox{\rm diag}
(1, 4 \lambda_2) \,.}
Parametric equations, $p=\phi(\lambda)$, for the stratum can be
written in the following form:
\begin{eqnarray} \label{parametrizzazione2}
\phi_1(\lambda) &=& {\lambda_1}^2 + \lambda_2  \,,\nonumber\\
\phi_2(\lambda) &=& \lambda_2\,,\nonumber\\
\phi_3(\lambda) &=& -2 \, \sqrt{3}\,{\lambda_1}^{3}\,,\nonumber\\
\phi_4(\lambda) &=& - 2 \,\sqrt{3} \, \lambda_1 \lambda_2 \,, \nonumber\\
\phi_5(\lambda) &=& 4 {\lambda_1}^{2} \lambda_2 \,,
\end{eqnarray}
for $\lambda$ in the range $\Delta = \left \{(\lambda_1,\lambda_2)
\in \real^2 \;|\; \lambda_2 >0 \right \}$. The rank of
$J(\lambda)$   is equal to 2 on the whole $\Delta$.

\subsubsection{Stratum $S^{(2)}_B$.}
A typical point
is $x_{\rm t}=(1,1,0,0,0,0,0,0)$ and the isotropy subgroup $H$ of
$G$ at $x_{\rm t}$ is the subgroup
$\integer_2\times\integer_2\times\integer_2$ formed by the
reflections of the axes in $\real^3$.
The vector space left invariant by $H$ reduces to
\begin{displaymath}
V = \left\{x\in \real^8 \mid
x_3=x_4=x_5=x_6=x_7=x_8=0\right\}\,.
\end{displaymath}
The group $K$ is the finite group of order 6 generated by
reflections with root system of type A$_2$ \cite{fini} with
generators

\begin{displaymath}
\left( \begin{array}{cc}
 -1/2 & \sqrt{3}/2 \\
\sqrt{3}/2 & 1/2
\end{array} \right) \,,\quad \quad
\left( \begin{array}{cc}
 -1/2 & -\sqrt{3}/2 \\
\sqrt{3}/2 & -1/2
\end{array} \right)\;.
\end{displaymath}
A MIB for $\mbox{\rm Stab }(H,G)$ is
\begin{equation}
(\lambda_1,\lambda_2) = \left( {x_1}^2 + {x_2}^2\,,\,
   x_1 ( {x_1}^2 - 3 \,{x_2}^2) \right)\,.
\end{equation}
The associated matrix $\widehat\Lambda(\lambda)$ turns out to be
\eq{\widehat\Lambda(\lambda)=\left(\begin{array}{ll}
4 \lambda_1 & 6 \lambda_2 \\
6 \lambda_2  & 9 {\lambda_1}^2
 \end{array}\right)\,.}
Parametric equations, $p=\phi(\lambda)$, for the stratum can be
written in the following form:
\begin{equation} \label{parametrizzazione3}
\left(\phi_1(\lambda),\phi_2(\lambda),\phi_3(\lambda),
\phi_4(\lambda), \phi_5(\lambda) \right)= \left(\lambda_1  \,,
  0\,, - 2\, \sqrt{3} \lambda_2\,, 0 \,, 0 \right)\,,
\end{equation}
for $\lambda$ in the range $\Delta = \left \{(\lambda_1,\lambda_2)
\in \real^2 \;|\; \lambda_1 >0 \; \mbox{\rm and} \; {\lambda_1}^3-
{\lambda_2}^2>0 \right \}$. The rank of $J(\lambda)$  is equal to
$2$ for all $\lambda\in\Delta$.

\subsubsection{Stratum $S^{(1)}$.}
 A typical point is
$x_{\rm t}=(1,0,0,0,0,0,0,0)$ and the isotropy subgroup $H$ of $G$
at $x_{\rm t}$ consists of the O(3) block-diagonal matrices of the
form diag$(\pm 1,O)$, with $O \in $ O(2). The vector space left
invariant by $H$ reduces to
\begin{displaymath}
V = \left\{x\in \real^8 \mid
x_2=x_3=x_4=x_5=x_6=x_7=x_8=0\right\}\,,
\end{displaymath}
the group $K$ is trivial and a MIB for $\mbox{\rm Stab }(K,V)$ is
simply $\lambda_1 =  x_1$. The associated matrix
$\widehat\Lambda(\lambda)$ reduces to $1$.

Parametric equations, $p=\phi(\lambda)$, for the stratum can be
written in the following form:
\begin{equation} \label{parametrizzazione4}
\left(\phi_1(\lambda),\phi_2(\lambda),\phi_3(\lambda),\phi_4(\lambda),\phi_5(\lambda)\right) =
\left( {\lambda_1}^2  \,, 0\,, -2 \, \sqrt{3}\,{\lambda_1}^3\,,
  0 \,,  0 \right)\,,
\end{equation}
for $\lambda$ in the range $\Delta = \real$.
One can verify that the rank of $J(\lambda)$  diminishes
in the allowed range $\Delta$ for $\lambda_1=0$.

\subsection{A physically interesting non coregular variant of the example}
The example presented in the first part of this section has been
suggested by a paper of Longa and Trebin \cite{1}, devoted to the
construction of a phenomenological theory of polar structures in
chiral biaxial liquid crystals, through the exploitation of the
properties, under transformations of a symmetry group SO(3), of a
symmetric and traceless tensor order parameter field $Q$ and of a
polar vector field $\bi{P}$.

In fact, the situation examined by Longa and Trebin can be
recovered from the example worked about, considering the action in
$\real^8$ of the subgroup SO(3) of O(3). In this case, we obtain a
non-coregular linear group $G'$. To get a MIB for $G'$, a new
$G'$-invariant polynomial $p_6(x)$, of degree 6, has to be added
to the MIB (\ref{BaseInt}):
\begin{equation}
p_6 = 2\, \sqrt{2} \left(\bi{P} \wedge Q \,\bi{P}\right) \cdot
({Q}^2 \bi{P})\;.
\end{equation}
The \wPm\/ associated to the enlarged MIB can be constructed from the
one shown in (\ref{BaseInt}) by adding one more row and
column:
\begin{eqnarray}
\widehat P_{1\,6}&=& 12 p_6\,, \nonumber\\
\widehat P_{2\,6}&=& 6 p_6\,, \nonumber\\
\widehat P_{3\,6}&=& 0\,, \nonumber\\
\widehat P_{4\,6}&=& 0\,, \nonumber\\
\widehat P_{5\,6}&=& 12\, p_1 p_6\,, \nonumber\\
\widehat P_{6\,6}&=& -\frac{1}{9}\left[ -{{p_1}}^2\,{{p_4}}^2   +
    {p_4}\,\left( {p_3} + {p_4} \right) \,{p_5} - 3\,{p_1}\,{{p_5}}^2
    +
    2\,{p_1}\,{p_2}\,\left( - {p_3}\,{p_4}   - {{p_4}}^2 +
       6\,{p_1}\,{p_5} \right) + \right. \nonumber\\
       && \left.\hspace{2em}-
    {{p_2}}^2\,\left( {{p_3}}^2 - 2\,{p_3}\,{p_4} - 3\,{{p_4}}^2 +
       12\,{p_1}\,{p_5} \right) \right]\,.
\end{eqnarray}
The added element $p_6(x)$ is a ``numerator invariant'' for the
Molien function \cite{Stan} and is, therefore, algebraically dependent on the
other elements of the MIB. The relevant relation can be easily
obtained from one of the irreducible polynomial factors of the
enlarged $\widehat P$ matrix:

\begin{equation}
243 \,{p_6}^2 + A(p) =0\,,
\end{equation}
where $A(p)$ is the same (see (\ref{attivo})) as in the coregular case.

Being $G'$ a subgroup of $G$, the lattice of isotropy types of
(SO(3),$\real^8$) is easily found from the following relation,
holding true for all $x\in \real^8$:

\eqlX{G'_x\, =\, G_x \bigcap G'\;.}{isotropy}

An explicit calculation shows that, the number of strata of
(SO(3),$\real^8$) is lower than the number of strata of
(O(3),$\real^8$). In fact, the four--dimensional stratum $S^{(4)}$
of the second group space is part of the principal stratum of
(SO(3),$\real^8$). The result is summarized in \tref{tab4}, where
the notations are the same used for the generators in the O(3)
case.

As explained at the end of section 2, if SO(3) is assumed to be
the (largest) symmetry group of the free energy, the strata of
$\real^8/G'$ are in a one-to-one correspondence with the
structural phases of the system, {\em i.e.}, with the phases of
the system that are identified only on the basis of their SO(3)
symmetry. An easy comparison with the results summarized in
table~I (p. 3460) of \cite{1}, shows that distinct polar states
($F_{\mbox{\rm ch}}$, $F_{Bg}$ and $F_{U2}$) defined by Longa and
Trebin, lie on a same stratum (the principal stratum) of
$(\mbox{\rm SO}(3),\real^8)$,
 meaning that their
symmetry (orbit type) is the same: they form a unique structural
phase. As stated above, the number of structural phases is
increased
 if the symmetry group of the free energy is enlarged to
O(3) (see our \tref{tab5}), but also in this case the polar states
$F_{Bg}$ and $F_{U2}$ lie on the same stratum $S^{(4)}$.

In their paper, Longa and Trebin make a big effort to determine
the range of the orbit map and produce a classification of the
polar states that, however, appears to be not based on their
symmetry properties with respect to the symmetry group of the free
energy. The effectiveness and mathematical rigor of the orbit
space approach to the study of invariant functions clearly emerges
from a comparison with the easy calculation, that led us,
essentially, to more rigorous results, and with the additional
important advantage of keeping, at every step, a clear
correspondence between geometrical structures (strata) and
symmetry of the corresponding physical configurations. In fact,
the parametrization technique becomes very important for orbit
spaces of group actions where more than three invariants are
involved; here the geometrical intuition based on drawing the
shape of the orbit space loses most of its efficiency.

\ack{This paper is partially supported by INFN and MIURST (40\%
and 60\%).}


\Tables

\begin{table}
\caption{\label{tabsing} Relations defining strata of dimension
$<4$ in the orbit space of (O(3),$\real^8$).}
\begin{center}
\begin{tabular}{ccc}
\br
&&\\
Stratum & Equalities & Inequalities \\
&&\\
\mr
&&\\
$S^{(1)}$ & $p_2=p_4=p_5=12 \, {p_1}^3-{p_3}^2   = 0$ & $p_1>0$ \\
&&\\
\hline
&&\\
$S^{(2)}_{A}$ & $p_1 - p_2- {\displaystyle \frac{{p_4}^2}{12 \,
{p_2}^{2}}}= p_3 - {\displaystyle\frac{{p_4}^3}{12 \, {p_2}^{3}}}=
 p_5-  {\displaystyle\frac{{p_4}^2}{3 \, p_2}}=0$ & $p_2>0$ \\
&&\\
\hline
&&\\
$S^{(2)}_{B}$ & $p_2=p_4=p_5= 0$ & $12 \, {p_1}^3-{p_3}^2>0$ \\
&&\\
\hline
&&\\
$S^{(3)}$ & $p_3-p_4 \left( 3 - {\displaystyle \frac{3 p_1}{p_2}}
+ {\displaystyle
                             \frac{{p_4}^2}{3 {p_2}^3}}\right)=
                   p_5-  {\displaystyle \frac{{p_4}^2}{3 \, p_2}}=0$ & $p_1>p_2>0\,,\;\;$\\
           &     &
                   $12\, p_1 {p_2}^2- 12\, {p_2}^3 - {p_4}^2 >0 $\\
                   &&\\
                  \br
\end{tabular}
\end{center}
\end{table}

\begin{table}
\caption{ \label{tab2} Isotropy type stratification of the orbit
space $V/K$, whose principal stratum is diffeomorphic to the
subprincipal stratum $\widehat S^{(4)}$ of the orbit space
$\real^8/G$.}
\begin{center}
\begin{tabular}{ccc}
\br
&&\\
Stratum & Equalities & Inequalities \\
&&\\
\mr
&&\\
$\Sigma^{(2)}$ & $\lambda_3=\lambda_4= 0$ & $\lambda_2>0$ \\
&&\\
\hline
&&\\
$\Sigma^{(3)}$ & $ 4 \lambda_2  {\lambda_3}^{2}= {\lambda_4}^2$ &
${\lambda_3}>0$ and
$\lambda_2 \geq 0$ \\
&&\\
\hline
&&\\
$\Sigma_p$ &    & $\lambda_2>0$ and $\lambda_3>0$ and
$ 4 \lambda_2  {\lambda_3}^{2}- {\lambda_4}^2 >0$ \\
&&\\
\br
\end{tabular}
\end{center}
\end{table}

\begin{table}
\caption{\label{tab4} Strata and isotropy types of the space
(SO$(3),\real^8)$. For each orbit type, a typical point and the
corresponding isotropy subgroup (or its generators) are specified,
with reference to the form of the matrices $O$ appearing in
(\ref{mu2}); $O_{\pm}$ denote elements of O(2) with determinant
$\pm 1$, respectively.}
\begin{center}
\begin{tabular}{ccc}
\br
&&\\
Stratum & typical point & isotropy subgroup or set of generators \\
&&\\
\mr
&&\\
$S^{(5)}$ & (1,1,0,0,0,0,1,1,1) & $\{ \mbox{ e } \}$\\
&&\\
\hline
&&\\
$S^{(3)}$ & (1,1,0,0,0,0,1,0,0) & $\{\mbox{\rm diag}(1,-1,-1)\}$\\
        \hline
&&\\
$S^{(2)}_{A}$ & (1,0,0,0,0,0,1,0,0) & $\mbox{\rm diag}(1,O_{+}) $\\
&&\\
\hline
&&\\
$S^{(2)}_{B}$ & (1,1,0,0,0,0,0,0,0) & $\{\mbox{\rm diag}(1,-1,-1),\mbox{\rm diag}(-1,1,-1)\}$\\
&&\\
\hline
&&\\
$S^{(1)}$  & (1,0,0,0,0,0,0,0,0) & $\mbox{\rm diag}(1,O_{+})\cup\mbox{\rm diag}(-1,O_{-})$\\
&&\\
\br
\end{tabular}
\end{center}
\end{table}

\begin{table}
\caption{ \label{tab5} Correspondence between Trebin and Longa's
polar states and isotropy type strata for the group O(3); the
Graphical Representation (G.R.) of the  {\em typical point} is the
same as in \cite{1}: the straight lines denote the non degenerate
eigenvector directions of $Q$, whilst the arrows represent the
orientation of the polar vector $\bi{P}$ with respect to the
eigenvectors of $Q$. The state $F_{U2}$ of \cite{1}, for which the
angle between the polar vector $\bi{P}$ and the unique
non-degenerate eigenvector of $Q$ is different from $0$ and
$\pi/2$, actually belongs to the stratum $S^{(4)}$. Therefore, it
is written in parenthesis, since it is equivalent (from the
symmetry point of view) to the state $F_{Bg}$.}
\begin{center}
\begin{tabular}[c]{cccc}
\br
Stratum & isotropy subgroup   &   G. R. of the {\em typical point}   & Phase in \cite{1}\\
\mr
$S^{(5)}$  &  $\{ \mbox{\rm e} \}$ &
\begin{texdraw} \drawdim mm
\linewd 0.3 \move(0 0) \lvec(9.375 9.375) \lvec(28.125 9.375)
\move(9.375 9.375) \lvec(9.375 23.75) \move(9.375 9.375)
\avec(18.75 15.625) \linewd 0.01 \lvec(18.75 1.5625) \lvec(1.5625
1.5625) \move(18.75 1.5625) \lvec(26.5625 9.375) \move(18.75
15.625) \lvec(9.375 23.4375) \move(9.375 9.375) \lvec(18.75
1.5625)
 \end{texdraw}
 & $F_{\rm ch}$\\
&&&\\
\hline
&&&\\
$S^{(4)}$ & $\{$diag$(\pm 1,1,1)\}$ & \begin{texdraw} \drawdim mm
\linewd 0.3 \move(0 0) \lvec(9.375 9.375) \lvec(28.125 9.375)
\move(9.375 9.375) \lvec(9.375 23.75) \move(9.375 9.375)
\avec(18.75 20.3125) \linewd 0.1 \move(18.75 20.3125) \lvec(18.75
9.375) \move(18.75 20.3125) \lvec(9.375 20.3125)
 \end{texdraw}& $F_{Bg}$ ($F_{U2}$)\\
&&&\\
\hline
&&&\\
$S^{(3)}$ & $\{$diag$(1,\pm 1,\pm 1)\}$ &\begin{texdraw} \drawdim
mm \linewd 0.3 \move(0 0) \lvec(9.375 9.375) \lvec(28.125 9.375)
\move(9.375 9.375) \lvec(9.375 23.75) \move(18.75 4.6875)
\avec(18.75 20.3125)
 \end{texdraw}& $F_{B||}$ \\
&&&\\
\hline
&&&\\
$S^{(2)}_{A}$ & $\{$diag$(1,O)$,  $ O \in $ O$(2)\}$
&\begin{texdraw} \drawdim mm \linewd 0.3 \move(9.375 9.375)
\lvec(9.375 23.75) \move(18.75 4.6875) \avec(18.75 20.3125)
 \end{texdraw}& $F_{U1}$  \\
&&&\\
\hline
&&&\\
$S^{(2)}_{B}$ & $\{$diag$(\pm 1,\pm 1,\pm 1)\}$ & \begin{texdraw}
\drawdim mm \linewd 0.3 \move(0 0) \lvec(9.375 9.375) \lvec(28.125
9.375) \move(9.375 9.375) \lvec(9.375 23.75)
 \end{texdraw}&  \\
&&&\\
\hline
&&&\\
$S^{(1)}$ & $\{$diag$(\pm 1,O)$, $ O \in $ O$(2)\}$
&\begin{texdraw} \drawdim mm \linewd 0.3 \move(9.375 9.375)
\lvec(9.375 23.75) \setgray 1 \lvec(10.0 23.75)
 \end{texdraw}&  \\
\br
\end{tabular}
\end{center}
\end{table}

\Figures

\end{document}